\documentclass[aps,twocolumn,amsmath,amssymb]{revtex4}

\usepackage{hyperref}
\usepackage{graphicx}

\begin{document}

\newcommand{\Det}{\,{\rm{Det}\,}}
\newcommand{\Tr}{\,{\rm{Tr}\,}}
\newcommand{\tr}{\,{\rm{tr}\,}}

\newcommand{\bomega}{\bar{\Omega}}
\newcommand{\wsf}{\omega_{\rm sf}}
\newcommand{\sign}{{\rm sign}}

\renewcommand{\Re}{{\rm Re}}
\renewcommand{\Im}{{\rm Im}}

\let\oldmarginpar\marginpar
\renewcommand\marginpar[1]{\-\oldmarginpar[\raggedleft\small\sf #1]{\raggedright\small\sf #1}}

\newcommand{\COMMENT}{\marginpar}
\newcommand{\FigWidth}{\columnwidth}

\title{Self-generated locality near a ferromagnetic quantum-critical point}

\author{A. V. Chubukov}
\affiliation{Department of Physics, University of Maryland,
College Park, MD, 20814-4111}


\begin{abstract}
We analyze the behavior of interacting fermions near a ferromagnetic 
 Stoner instability.
 We show that the Landau damping of the spin susceptibility is a relevant perturbation near a ferromagnetic quantum-critical point (FQCP). We argue that,
 as the system 
 approaches a FQCP, 
 the fermionic self-energy
 crosses over from  predominantly momentum dependent 
away from the transition to predominantly frequency dependent in the immediate vicinity of the transition. We argue that due to this self-generated locality, 
 the quasiparticle effective mass does not diverge before a FQCP is reached. 
\end{abstract}


\maketitle

\section{Introduction}

Over 50 years, Fermi liquid theory serves as the
 basis for our understanding of the behavior of electrons in metals~
\cite{1,2,3}. Developed by Landau to describe the behavior of $^3He$ at low $T$~\cite{1} 
 and based on a minimal number of postulates, it allows to describe the 
behavior of interacting fermions not only qualitatively, but also 
quantitatively~(see e.g.,~\cite{new_Greywall}).

The key postulate of  Fermi liquid theory is  that
 the interaction  between fermions with energies near the Fermi energy 
 is weak in any Fermi liquid, and does
 not qualitatively change the structure of low-energy fermionic states
 compared to a Fermi gas.
In the Green's function language, the condition of weak scattering 
 near the Fermi surface implies that the fermionic self-energy 
 $\Sigma (k \omega)$ is linear in $\omega$ and in $k-k_F$ 
at smallest  frequencies and  smallest $|k-k_F|$. 
 In particular, at $k=k_F$, 
\begin{equation}
\Sigma(k_F, \omega) = \lambda \omega
\label{1}
\end{equation}
The imaginary part of $\Sigma (\omega)$,  responsible for the scattering from one electronic state into the other, and hence, for a finite lifetime of a given fermionic state, must be smaller than $\omega$, i.e., should behave 
at low frequencies as 
\begin{equation}
\Sigma^{\prime \prime} (\omega) \propto \omega^{1+\alpha}, \alpha >0    
\label{2}
\end{equation}
The conditions specified by 
(\ref{1}) and (\ref{2}) imply that at low energies, the dominant effect of electron-electron interaction is the shift of the energy levels, proportional to the deviation of the energy level from the Fermi surface. 
At the same time, the levels themselves
 remain intact, i.e., they are not destroyed by the interaction. 
 Eqs. (\ref{1}) and (\ref{2}) 
imply that in the limit of small $\omega$ and small $|k|-k_F$, the Green's function for interacting fermions has the same form as for non-interacting fermions, modulo the renormalization of the Fermi velocity and the overall 
 $Z$ factor: 
\begin{equation}
G(k,\omega) = \frac{Z}{\omega - v^*_F (|k|-k_F) + i \delta {\text sgn} \omega}
\label{3}
\end{equation}
where $v^*_F = k_F/m^*$, and $m^*$ is the effective mass.

Since the corrections due to electron-electron interactions do not change the functional form of the fermionic propagator, the thermodynamic characteristics 
 of a Fermi system, which describe  the system's reaction to small external
 perturbations such as temperature or magnetic field, must retain the
 same functional form as for free fermions. This, in particular, implies
 that the specific heat is linear in $T$ at low temperatures,
$C(T) \propto T$,
and the static spin susceptibility reduces to a constant at $T=0$.
The linear in $T$ specific heat and the constant spin susceptibility are
 the two fundamental properties of a Fermi liquid in any $d>1$.

In this communication, we consider the situation when the system at $T=0$ 
is close  to a density-wave instability at  $q=0$. The 
most straightforward example is a system 
 near a ferromagnetic quantum critical point (FQCP). 
 Spin fluctuations can be either isotropic or
 anisotropic, i.e., of Ising type, due to spin-orbit coupling. 
 Our conclusions are applicable for both cases. For definiteness,
 we consider isotropic case. 
We assume that the system is weakly coupled far away from the transition. 
This implies that the corrections to a Fermi liquid  are generally 
small. Near the transition, however, the bosonic mode that mediates interaction between fermions becomes soft, and 
 self-energy corrections  increase~\cite{hertz,doniach,sachdev,hf-th,advances,meta,bedell,sc_qcp}.
 In $d<3$, the dimensionless coupling constant $\lambda$ scales as a positive 
 power of the ferromagnetic correlation length $\xi$:
$\lambda = g/W~(\xi/a)^{3-d}$, where $g$ is the coupling, $a$ is the 
 interatomic spacing, and $W = v_F/a$ is of order of the fermionic bandwidth.
 At $\xi\sim a$, $\lambda \sim g/W$ is
 small if, as we assume,  the interaction is small compared to $W$. 
However, as $\xi$ increases, $\lambda$ also increases and diverges at a FQCP,
 where $\xi = \infty$. 
 
The issue brought about
 in several recent publications~\cite{khodel,yak_khodel} 
 is whether in this situation
 a Fermi liquid survives right up to a critical point, or it is destroyed
 already at some distance away from criticality. The possibility that a Fermi 
 liquid is destroyed before a FQCP is based on the observation~\cite{yak_khodel}
 that if
  the fully renormalized  interaction is static, then
  the effective mass $m^*/m = 1/(1-\lambda)$ diverges already at $\lambda =1$, i.e., at some distance from a FQCP. It was suggested~\cite{khodel}
 that a new intermediate phase termed as ``fermionic condensate'' 
may emerge near a FQCP. 

We will show that the  $1/(1-\lambda)-$ dependence of the effective 
  mass holds only as long as the
 retardation of the interaction is a small perturbation, and the self-energy
predominantly depends on $k$: $\Sigma (k, \omega) \approx \Sigma (k)$.
 We show that, as $\xi$ increases, the
 self-energy crosses over from $\Sigma (k)$ to $\Sigma (\omega)$ already at $\lambda \sim (g/W)^{(d-1)/2} \ll 1$.  Above the crossover,
 the expression for the mass changes from $m^*/m = 
1/(1 - \lambda)$ to $m^*/m = 1 + \lambda$. 
Then, as the system enters into the strong coupling regime $\lambda > 1$, 
 the effective mass increases, but it does not diverge until the system reaches  a FQCP.
As a result, no new intermediate phase phase appears near FQCP.

A similar situation has recently been found in ~\cite{cgy}
 for a transition in a spatially isotropic  system into
 a density-wave state at at finite $q=q_0$. 
There is a qualitative similarity between the cases $q=q_0$ considered in ~\cite{cgy} and $q=0$ considered here --in both cases the effective 
mass does not diverge before FQCP. However, the physics of the
 crossover is somewhat different in the two cases (see below).

In our analysis  we neglect two other peculiarities of the
 system behavior near a FQCP. The first is
 the non-analytic momentum dependence of the 
 static spin susceptibility~\cite{belitz}. It either makes the
 ferromagnetic transition first order, or leads to spiral distortion of a second-order ferromagnetic ordering.
The second is the pairing instability~\cite{bedell,sc_qcp,lonzarich}.
 It leads to the development of a dome 
 on top of the FQCP where  the system becomes 
a $p$-wave superconductor. These two peculiarities compete with each other, but both affect the system behavior in the immediate vicinity of the transition, when the locality is already self-generated. In this paper
 we will be primarily interested in the origin for
 the locality, and how the self-energy 
 crosses over from $\Sigma (k)$ to $\Sigma (\omega)$. 

\section{The self-energy in a Fermi liquid near a FQCP}

\subsection{The spin-fermion model}

\begin{figure}
\includegraphics[width=\FigWidth]{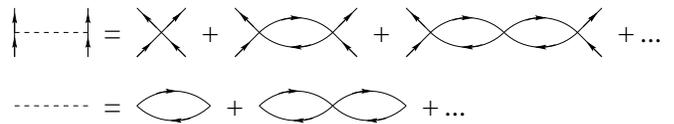}
\caption{\label{fig_RPA} 
The schematic derivation of the effective interaction within the RPA
 approximation. The dashed line is the effective bosonic propagator. 
 Near SDW instability, the vertices between solid and dashed lines contain
 spin $\sigma$ matrices.}
\end{figure}

Following earlier studies~\cite{hertz} we 
assume that the system behavior near a FQCP is 
described by the spin-fermion model, in which fermion-fermion interaction is mediated by collective spin fluctuations peaked at $q=0$:
\begin{equation}
\label{Hint}
  {\cal H}_{\rm int} = - \sum_{\bf{q},\bf{k},\bf{k}'}
  \psi_{\bf{k}+\bf{q},\alpha}^{\dagger} \psi_{\bf{k}'-\bf{q},\beta}^{\dagger} 
  V (\bf{q}) {\vec \sigma}_{\alpha,\gamma} {\vec \sigma}_{\beta,\delta}
\, \psi_{\bf{k}',\gamma} \psi_{\bf{k},\delta} .
\end{equation}
where  $V({\bf q})\approx (g/a^2)/(\xi^{-2}+q^2)$. This model describes the 
 physics at low energies and can, in principle, be obtained  
  from the original Hubbard-type model of short-range 
electron-electron interaction on a lattice, after fermions with  energies comparable to $W$ are integrated out~\cite{advances}. Within RPA approximation, this procedure is illustrated in Fig.\ref{fig_RPA}.
 This separation of scales into a 
low-energy part and a high-energy part makes sense only when $g <W$ as we 
will see below that quantum-critical behavior 
 extends to  energies of order $\omega_0 \sim g^2/W$. 
 The collective spin fluctuations are
 low-energy degrees of freedom near FQCP, and they remain in the theory after
 the high-energy degrees of freedom are eliminated. For  fermions,
the restriction to low-energies implies that their dispersion can be linearized near the Fermi surface: $\epsilon_k = v_F (k-k_F)$. 

To keep $g$ smaller than $W$ and, at the same time,
 bring the system to a ferromagnetic instability requires extra assumptions.
  Within the RPA approximation~\cite{scal}, 
the full spin susceptibility $\chi (q) = \chi_0 (q)/( 1 -g \chi_0 (q))$, 
 where $\chi_0 (q) \sim 1/W$ is the static 
 spin susceptibility of free fermions.
 Ferromagnetic instability then 
occurs only when $g \sim W$. There are two ways out of this. 
First,  if the range of the  
interaction $r_0$ is 
 much larger than the interatomic spacing (i.e., $r_0 \gg a$), the 
 condition for FQCP becomes $g /W \sim a/r_0 \ll 1$, i.e., 
the smallness of $g/W$ is consistent with the closeness to FQCP~\cite{gorkov}.
Second,  one can use the fact that the renormalizations outside RPA 
 make effective coupling frequency dependent $g = g(\omega)$. 
The ferromagnetic instability
 is then determined by $g (\omega \sim W) \sim W$, while the coupling constant $\lambda$ involves $g(\omega)$ at small $\omega << W$. 
 The separation of scales  then becomes 
 possible if $g(\omega << W) << g(W)$. 
We assume that either the range of the interaction,
 or its frequency dependence validate the condition 
\begin{equation}
g (\omega =0)/W \ll 1
\end{equation}
 near  a FQCP. 

\subsection{Fermionic self-energy} 
\label{se}
 
\begin{figure}
\includegraphics[width=\FigWidth]{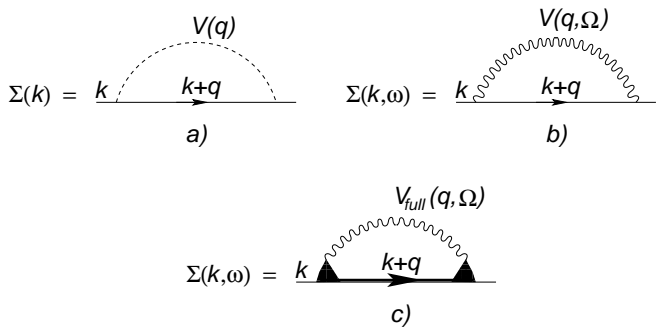}
\caption{\label{fig_1} 
(a,b) -- The self-energy diagram with the bare interaction $V(q)$, and
 the effective interaction $V (q, \Omega)$. (c) -- The diagram for the 
full self-energy. It contains full fermionic propagator, full vertices, and 
 the fully renormalized $V_{full} (q,\Omega)$. }
\end{figure}

Consider first the self-energy within
 the perturbation theory for Eq. (\ref{Hint}).
The self-energy diagram to first order in $V(q)$ is presented in Fig.
 \ref{fig_1}a.
In analytic form,  at $T=0$   
\begin{equation}
\Sigma ({\bf k}, \omega_m) = 3 \int \frac{d^dq d\Omega_m}{(2\pi)^{d+1}}~
 V({\bf q}) ~G^{(0)}_{{\bf k+q}, \omega_m + \Omega_m}
\label{1.7}
\end{equation}
where $G^{(0)}_{{\bf k}, \omega_m} = (i\omega_m -\epsilon_k)^{-1}$ and
$\epsilon_k = v_F (k-k_F)$. 
The full Green's function is related to $\Sigma$ as $G^{-1} (k,\omega_m) = 
(G^{(0)} (k, \omega_m))^{-1} + \Sigma (k, \omega_m)$. Since we are interested in finite $\omega_m$ and finite deviations from the Fermi surface, it is convenient to subtract the constant  $\Sigma(k_F,0)$ from the self-energy in (\ref{1.7}) 
 before evaluating the integral. This constant $\Sigma(k_F,0)$ 
accounts for the renormalization of the chemical potential and does not affect the physics near FQCP. We will just neglect it below.
Expanding the fermionic dispersion $\epsilon_{{\bf k+q}}$ at small $q$ as
$\epsilon_{{\bf k+q}} = \epsilon_k + v_F q \cos \theta$, where $\theta$ is the angle between ${\bf k}$ and ${\bf q}$, we obtain
\begin{equation}
\Sigma ({\bf k}, \omega_m)  = (i\omega_m - v_F \delta k) I (\delta k, \omega_m)\label{1.10}
\end{equation}
where $\delta k = k-k_F$, and 
\begin{eqnarray}
&&I (\delta k, \omega_m) = -3 \int \frac{d^dq d\Omega_m}{(2\pi)^{d+1}}~
 V({\bf q}) \nonumber \\
&& \frac{1}{[\Omega +i v_F q \cos \theta][\omega + \Omega +i v_F (\delta k + q \cos \theta)]}. 
\label{1.11}
\end{eqnarray} 
We see from (\ref{1.10}) that if $I(\delta k, \omega_m)$ is non-singular
 at small $\omega_m$ and $\delta k$, the self-energy scales as $\omega_m$
 and $\delta k$. This implies that the self-energy modifies the parameters in the Green's function but does not affect its functional form compared to free fermions, in 
 agreement with the key postulate of the Fermi liquid theory.

Although $I (\delta k, \omega_m)$ is non-singular, it has to be evaluated with extra care by two reasons. 
First, for a constant $V(q)$, the integrand in (\ref{1.11}) formally 
diverges at large $\Omega_m$ and $q$, hence the ordering of the integrations over momentum and frequency is relevant. Second,
 the integrand in (\ref{1.11}) contains two  poles separated
 only by $\omega$ and $v_F \delta k$. In this situation, one cannot just set 
$\omega_m = \delta k =0$ in evaluating $I (\delta k, \omega_m)$, but rather should 
 consider the limit $\omega,~\delta k \rightarrow 0$. 
A similar situation occurs near a QCP in which the instability occurs 
 at a finite $q$~\cite{cgy}. 
To evaluate the integral at finite $\delta k$ and $\omega_m$, 
we note that $V(q)$ can be approximated by a constant only at small $q$,
 while at large momenta, $V(q)$ eventually vanishes for any realistic potential. The vanishing of $V(q)$  at large $q$
  provides the ultraviolet regularization of the integral in (\ref{1.11}). 
 Since the regularization is provided by the momentum dependence,
 one should first integrate over frequency in infinite limits
 (the corresponding integral in (\ref{1.11}) is convergent), 
 and then integrate over ${\bf q}$.
  For practical purposes, it is convenient to introduce
$q \cos \theta = q_1$. Then $\int d^d q = \int d^{d-1} q_{\perp}
 d q_1$. Performing the  integration over $\Omega$,
 we find that the frequency integral is nonzero when 
 $q_1$ and $q_1 + \delta k$ have different signs, and the two poles in
(\ref{1.11}) are located in different half-planes of frequency. 
For $q_1$ within this interval, the 
 integration over frequency yields
\begin{eqnarray}
&&\int \frac{d\Omega}{2\pi} \frac{1}{[\Omega +i v_F q_1][\omega + \Omega +i v_F (\delta k + q_1)]} \nonumber \\
&&=  - \frac{1}{i\omega_m - v_F \delta k}
\label{1.13}
\end{eqnarray}
  For other values of $q_1$, the poles are in the same half-plane of frequency,
 and the frequency integral vanishes.
 Since the result of the frequency integration in (\ref{1.13}) 
does not depend on  $q_1$, the subsequent
 integration over $q_1$  just gives  the length of 
the interval where the poles are in different half-planes of frequency, i.e.,
$\delta k$. As $\delta k$ is vanishingly small, 
 $V(q_{\perp},  q_1)$ can be safely replaced by
 $V(q_{\perp}, 0)$.  We then obtain
\begin{equation}
I (\delta k, \omega_m) = \lambda  \frac{v_F \delta k}{i \omega - v_F \delta k}
\label{1.16}
\end{equation}
where
\begin{equation}
\lambda = \frac{3}{v_F}  \int \frac{d^{d-1} q_{\perp}}{(2\pi)^d} V(q_{\perp})
\label{1.15}
\end{equation}
Accordingly,
\begin{equation}
\Sigma (k, \omega_m) = \lambda v_F \delta k
\label{1.17}
\end{equation} 
We see that the self-energy depends only on momentum. This result could 
indeed be anticipated as the interaction does not depend on frequency.
 One  could eliminate the frequency dependence in $\Sigma$ 
at the very first step of the calculations by 
shifting the frequency variable in Eq. (\ref{1.7}) 
 from $\Omega$ to $\omega + \Omega$.
 We will see, however,  that the integration procedure presented 
 above is more appropriate as it allows for straightforward
  extensions to the case when the interaction does depend on frequency.

 Substituting the self-energy from (\ref{1.17}) into the Green's function, we
 obtain $G^{-1} (k, \omega_m) = i \omega_m - \delta k (1- \lambda) = i\omega_m - v^*_F \delta k$, i.e.,
 the effective mass $m^* = p_F/v^*_F$ is
\begin{equation}
m^* =  \frac{m}{1-\lambda}
\label{1.18}
\end{equation}

By analyzing the computation of $m^*$, we see that the renormalization of the 
 fermionic mass indeed comes from intermediate fermionic states very near the Fermi surface ($\Omega \sim \omega,~v_F q_1 \sim \delta k$). The regions away from 
 the vicinity of the Fermi surface do not contribute to the mass 
 renormalization. This is again in agreement with the Fermi liquid theory.
 Note that from  the 
 mathematical point of view, the renormalization of the effective mass is a typical example of an ``anomaly''~\cite{anomaly}. Indeed,
 $I(0,0) =0$ because of the
 double pole in (\ref{1.11}), but the limit $\omega,\delta k \rightarrow 0$ of
 $I (\delta k, \omega_m)$ is finite because the double pole is splitted into two single poles, and 
 at  finite $\delta k$ there exists a range where these two
 poles are in different half-planes 
of frequency. The  smallness of the region where this happens 
is compensated by the  smallness of the distance between the 
 two poles. As a result, the momentum and frequency integration in Eq. (\ref{1.11}) yields $I \propto \delta k/(i\omega - v_F \delta k)$,
 which is finite at a nonzero $\delta k$ and 
depends on the ratio $\omega_m/(v_F \delta k)$. The same term $\delta k/(i\omega - v_F \delta k)$ is present in  the particle-hole bubble at small momentum and frequency transfer and accounts for the difference between $\Gamma^\omega$ and $\Gamma^k$ in the Fermi liquid theory~\cite{2,3}. 

The same type of analysis 
  can be performed for the spin susceptibility, and the result is that the renormalization of $\chi_s (Q=0, T=0)$ due to fermion-fermion interaction also comes from ``anomalous'' terms associated with the splitting of a double pole in the particle-hole bubble at small momentum and frequency transfer. From this perspective, Landau Fermi liquid theory can be viewed as 
 the  ``theory of anomalies''.  

Further, we see from (\ref{1.17}) that 
the  quasiparticle $Z$ factor is not renormalized by the self-energy and 
remains $Z=1$. Where the renormalization of $Z$ comes from? To answer this question, we return to Eq. (\ref{1.11}). The renormalization of $Z$ 
is  related to the renormalization of the frequency dependence of the Green's 
 function. The latter comes from a regular part in $I(\delta k, \omega_m)$, for which
${\text lim}_{\omega,\delta k \rightarrow 0} I (\delta k, \omega_m) = I(0,0)$.
  As we said, for a purely static $V(q)$, $I(0,0) =0$. Suppose now that the 
effective interaction acquires some  dependence on  frequency
 after high-energy fermions are integrated out. 
 This frequency dependence must be analytic, as relevant fermionic energies 
 are $O(W)$, i.e.,  are much larger than $\Omega$. 
Quite generally then,
\begin{equation}
V(q, \Omega) = \frac{g/a^2}{\xi^{-2}+q^2 + \left(\frac{\Omega}{v_s}\right)^2}
\label{1.21}
\end{equation}
where $v_s$ is the spin velocity. As the frequency dependence comes from fermions, $v_s \sim v_F$. 

For the interaction as in (\ref{1.21}), the frequency integral in 
Eq. (\ref{1.11}) for $I(0,0)$ has a double pole at $\Omega = i v_F q_1$, but
 also has two extra poles at $\Omega = \pm i v_s \sqrt{q^2 + \xi^{-2}}$. These two extra poles are present in both half-planes of frequency. Hence, even if the
 frequency integration is extended to the half-plane where there is no double pole, the result of frequency integration remains non-zero, and $I(0,0)$ becomes finite. 
As the two new poles are not associated with the splitting of a double pole,
 $I(0,0)$ is generally determined by fermionic states far away from the Fermi surface. 
In particular, at $\xi \sim a$, the poles in $V(q, \Omega)$ are located at
 $|\Omega| \sim v_s \xi^{-1} \sim v_F/a \sim W$.  By this reason,
 $Z$ is the input parameter for the Fermi-liquid theory, but it cannot be 
 evaluated within the Fermi liquid theory, which deals only with fermions 
 near the Fermi surface.

We see therefore that, in general, the second-order 
 self-energy $\Sigma (k, \omega)$ has 
 two contributions. The first, regular contribution,
 comes from fermionic states far from the Fermi surface
and yields
\begin{equation}
\Sigma_{reg} = (i\omega - v_F \delta k) I(0,0)
\label{1.22}
\end{equation}
The second, anomalous contribution, comes from fermionic states near the Fermi surface and yields
\begin{equation}
\Sigma_{an} = \lambda v_F \delta k
\label{1.22'}
\end{equation}
Substituting both terms into the expression for $G$, we see that the anomalous self-energy gives rise to the mass renormalization, while the 
regular part of the self-energy 
accounts for the renormalization of $Z$:
\begin{equation}
\frac{1}{Z} = 1 + I(0,0)
\label{1.23}
\end{equation}

\subsection {Arbitrary Fermi liquid}

The above consideration can be extended to an arbitrary strong interaction. 
The full self-energy diagram is presented in Fig. \ref{fig_1}b.
 It contains full Green's function, the fully renormalized interaction, and the full vertices. 
 As long as the full Green's function at low energies has the Fermi liquid 
form  of Eq. (\ref{3}), the self-energy has an anomalous term 
 which comes from the
 integration over electronic states very near the Fermi surface. 
For these contribution, the fully renormalized vertices can be approximated by their values at $\omega_m =0$ and $k =k_F$. The integration over $\Omega$ 
and over $q_1$ then proceeds the same way as before and yields 
\begin{equation}
\Sigma_{an} = \lambda Z v_F \delta k.
\label{1.25}
\end{equation}
where now 
\begin{equation}
\lambda = \frac{3}{v_F}  \int \frac{d^{d-1} q_{\perp}}{(2\pi)^d} {\bar V}(q_{\perp}) \Gamma^2 (q_{\perp})
\label{1.24}
\end{equation}
Here ${\bar V} (q_{\perp})=
{\bar V} (q_{\perp}, q_1 =0, \Omega_m =0)$ is the fully renormalized interaction, and $\Gamma (q_{\perp}) = \Gamma (q_{\perp}, q_1 =0, \Omega_m =0)$
 accounts for the vertex renormalization. 
Observe that $v_F$ in (\ref{1.25}) is a bare Fermi velocity. 
 
 The regular contribution can be
 rather complex because both ${\bar V}$ and $\Gamma$ depend on frequency.
Still, however, the regular self-energy depends only on $I(0,0)$ and is 
given by
\begin{equation}
\Sigma_{reg} = Z (i\omega - v^*_F \delta k) I(0,0).
\label{1.26}
\end{equation}
Substituting the full self-energy into the equation for the full Green's function, we obtain
\begin{eqnarray}
\frac{1}{Z} &=& 1 + I(0,0) \label{1.27} \\
\frac{1}{m^*} &=& \frac{1}{m}~ (1 - Z \lambda). 
\label{1.28}
\end{eqnarray}
Eq. (\ref{1.27}) determines $Z$, while Eq. (\ref{1.28}) determines the
 mass renormalization in terms of $Z$ and the coupling constant $\lambda$.

\subsection{The coupling constant vs $\xi$}

As  we discussed, the low-energy 
 theory based on the spin-fermion model 
is valid only
 when the spin-fermion interaction $g$ is much smaller than 
the fermionic bandwidth $W \sim v_F/a$. The dimensionless $\lambda$ scales with $g/W$,
 and is  a small number within the spin-fermion model when $\xi \sim a$.
Alternatively speaking, far away from quantum criticality, the spin-fermion model is only valid in the weak coupling limit.  
Consider, however, what happens with $\lambda$, given by (\ref{1.24})
 when the system approaches 
  FQCP, and $\xi$ diverges. Substituting $V(q) 
\approx (g/a^2)/(\xi^{-2}+q^2)$ into (\ref{1.24}), and assuming that the vertex renormalization is non-singular at small $q$, we obtain  
\begin{equation}
\lambda_d \propto \frac{\Gamma^2 g}{v_F}~\int \frac{dq q^{d-2}}{q^2 + \xi^{-2}} \label{2.2}
\end{equation}
For $d>3$, the integral over $q$ is infrared convergent, even when $\xi =\infty$. The integral is then determined by large $q \sim 1/a$ and yields
$\lambda \sim g/W$, that is, $\lambda$ remains small even at FQCP. 
For $d \leq 3$, the situation is, however, different. 
For $d =3$, the integral in (\ref{2.2}) depends logarithmically on the lower limit~\cite{bedell}, and 
\begin{equation}
\lambda_3 \propto \frac{\Gamma^2 g}{W}~\log{\frac{\xi}{a}}
\label{2.5}
\end{equation}
We see that, despite the smallness of the $g/W$ ratio, 
 the coupling constant diverges at FQCP. This implies that 
 in the immediate vicinity of FQCP, the weak coupling approximation becomes invalid. 

For $d<3$, the divergence of $\lambda$ becomes a power law: $\lambda_d \propto (\xi/a)^{3-d}$. For $d=2$ we have~\cite{advances,sc_qcp} 
\begin{equation}
\lambda_2 = \frac{3 \Gamma^2 g}{4 \pi W}~\frac{\xi}{a}
\label{2.6}
\end{equation}     
 
\subsection{The equivalence between diagrammatic technique and the Fermi liquid theory}

Before we proceed with the analysis of the crossover in the self-energy,
 it is instructive to compare the diagrammatic 
calculations of $m^*$ with  the Fermi liquid theory. 
For definiteness, we consider the $3d$ case.

 In the Fermi liquid theory, $m^*$ is related to the quasiparticle interaction function as
\begin{equation}
\frac{1}{m^*} = \frac{1}{m} - \frac{p_F}{2\pi^2} \int \frac{d \Omega}{4 \pi}
 f_c (\theta) \cos \theta 
\label{1.33}
\end{equation}
where $f_c$ is the charge component of the quasiparticle interaction function
$f_{\alpha,\gamma;\beta \delta} (p-p^\prime) = f_c (\theta)~ \delta_{\alpha \beta} \delta_{\gamma\delta} + f_s (\theta)~{\vec \sigma}_{\alpha\beta} {\vec \sigma}_{\gamma \delta}$, which also contains  the spin component $f_s$.
 $\theta$ is the angle between ${\bf p}$ and ${\bf p}^\prime$, which both are at the Fermi surface, $d \Omega$ is the solid angle.

\begin{figure}
\includegraphics[width=\FigWidth]{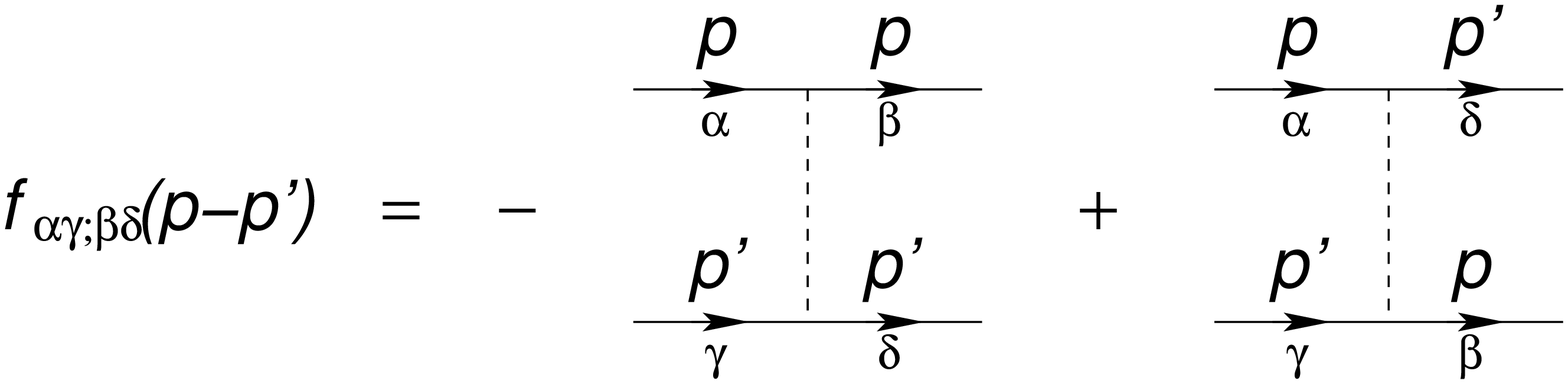}
\caption{\label{fig_2} 
The diagrams for the quasiparticle interaction function $f_{\alpha \gamma;\beta \delta} ({\bf p}, {\bf p}^\prime)$ to first order in the interaction.}
\end{figure}

To first order in the interaction,  the quasiparticle interaction function
 coincides,  up to the overall minus sign, with  the antisymmetrized 
interaction at zero momentum transfer. For spin-mediated interaction, we have
 (see Fig. \ref{fig_2})
\begin{equation}
f_{\alpha \gamma;\beta \delta} (p-p^\prime) = -V(0) {\vec \sigma}_{\alpha\beta} {\vec \sigma}_{\gamma \delta} + V(p-p^\prime) {\vec \sigma}_{\alpha\delta} {\vec \sigma}_{\gamma \beta}
\label{1.35_a}
\end{equation}
Using
\begin{eqnarray}
{\vec \sigma}_{\alpha\delta} {\vec \sigma}_{\gamma \beta} &=& -
\delta_{\alpha\delta} \delta_{\gamma \beta} + 2  \delta_{\alpha \beta} \delta_{\gamma\delta}, \nonumber \\
~\delta_{\alpha\delta} \delta_{\gamma \beta} &=& 
\frac{1}{2} \left({\vec \sigma}_{\alpha\beta} {\vec \sigma}_{\gamma \delta} +
 \delta_{\alpha \beta} \delta_{\gamma\delta}\right)
\label{1.35_b}
\end{eqnarray}
we obtain
\begin{equation}
f_c (\theta) = \frac{3}{2} V(\theta) 
\label{1.35_c}
\end{equation}
Substituting this into (\ref{1.33}), we obtain
\begin{equation}
\frac{1}{m^*} = \frac{1}{m} - \frac{3 p_F}{4\pi^2} \int \frac{d \Omega}{4 \pi}
 V (\theta) \cos \theta 
\label{1.33_a}
\end{equation}

This expression coincides with the lowest-order diagrammatic result. To see
 this, one can  evaluate the mass renormalization explicitly, using
 $V(q)$ from Eq. (\ref{Hint}) and $q = 2p_F \sin (\theta/2)$, and compare with 
 (\ref{1.15}), (\ref{1.18}). Alternatively, one can
 re-evaluate the diagrammatic expression to reproduce (\ref{1.33_a}). A way
 to do this is to integrate over frequency in (\ref{1.7}) without expanding $\epsilon_{k+q}$ in $q$. The frequency integration  yields
\begin{equation}
\Sigma (k, \omega) = \Sigma (k) = \frac{3}{2} ~\int \frac{d^3 q}{(2\pi)^3}~V(q)~ {\text sign}~ \epsilon_{k+q} 
\label{1.30}
\end{equation}
Subtracting a constant $\Sigma (k_F)$ from $\Sigma (k)$ and expanding 
the difference to first order in $\delta k =k-k_F$, we obtain
\begin{equation}
 \Sigma (k) = \frac{3}{2} v_F \delta k ~\int \frac{d^3 q}{(2\pi)^3}~\frac{\partial  ({\text sign}~ \epsilon_{k+q})}{\partial \epsilon_{k+q}} {\bf n}_{\bf k} {\bf n}_{{\bf k_F +q}} V (q)
\label{1.31}
\end{equation}
Using $\frac{\partial ( {\text sign}~ \epsilon_{k+q})}{\partial \epsilon_{k+q}}=
\delta(\epsilon_{k_F +q})$, introducing $q = 2k_F \sin (\theta/2)$ as both ${\bf k} = {\bf k}_F$ and ${\bf k}_F + {\bf q}$ are near the Fermi surface, 
 and shifting the  integration variable from 
 ${\bf q}$ to ${\bf k}_F + {\bf q}$, we obtain
 \begin{equation}
 \Sigma (k) = \frac{3 m k_F}{4\pi^2} v_F \delta k ~
\int \frac{d \Omega}{4 \pi}
 V (\theta) \cos \theta 
\label{1.31_a}  
\end{equation} 
Substituting this into the expression for the Green's function, we
 reproduce Eq. (\ref{1.33_a}). 

The extension of the analogy between the Fermi liquid theory and
diagrammatic technique to an arbitrary strong interaction requires some care.
 The diagrammatic theory yields, at arbitrary interaction strength,
\begin{equation}
\frac{1}{m^*} = \frac{1}{m} - \frac{3 Z p_F}{4\pi^2} \int \frac{d \Omega}{4 \pi}
 V (\theta) \cos \theta. 
\label{1.33_b}
\end{equation}   
The Fermi liquid formula, Eq. (\ref{1.33}) contains 
  $f_c (\theta)$ instead of $(3/2) Z V(\theta)$. In arbitrary Fermi liquid,
 $f_c$ is related to the charge component of the vertex function
$\Gamma^\omega$ as 
$f_c (\theta) = Z^2 \Gamma_c^\omega (\theta)$~\cite{2,3}.
 The full vertex $\Gamma^\omega$ includes all regular 
 vertex renormalizations, but does not include anomalous vertex corrections 
 from  the particle-hole bubble at small momentum and frequency transfer
 (the latter disappear in the limit $v_F \delta k/\omega \rightarrow 0$). 

\begin{figure}
\includegraphics[width=\FigWidth]{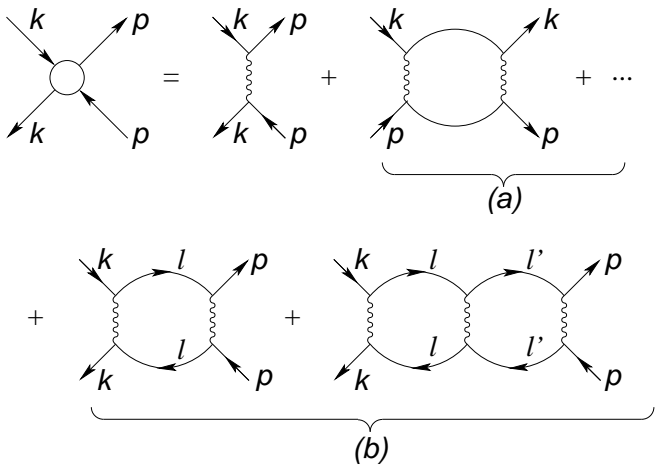}
\caption{\label{fig_3} 
The  diagrammatic series for $\Gamma^\omega$.
The diagrams  labeled as ``b'' must
 be included, but the diagrams labeled as ``a''
 should be excluded as their inclusion
 would amount to double counting.}
\end{figure}

We see that Fermi liquid and diagrammatic expressions for $m^*$ 
are equivalent 
 if  the relation between $V(\theta)$ and 
$f_c (\theta)$  is   $f_c (\theta) = (3/2) Z V (\theta)$, i.e., 
is the same as in (\ref{1.35_c}), but with one power of $Z$.
  To understand how this  $Z$ appears,
 consider the diagrammatic series for $\Gamma^\omega$ in terms
 of $V$, Fig.\ref{fig_3} 
The diagrams of Fig. \ref{fig_3}a should be neglected as their inclusion
 would amount to double counting - these renormalizations are already incorporated into the effective interaction $V (q, \omega)$,
 when high-energy fermions are integrated out in the RG-type procedure. The diagrams of Fig. \ref{fig_3}b, however,
 should be included into $\Gamma^\omega$ as in the RG treatment they  
vanish because of double poles.
Since $\Gamma^\omega$ does not include anomalous contributions from particle-hole bubbles, the  diagrams in Fig. \ref{fig_3}b are  non-zero only  
if $V(q, \omega)$ 
  depends on  frequency, otherwise the frequency integral of the convolution of two fermionic propagators  vanishes.  We remind 
that when the effective $V (q, \omega)$  depends on frequency, 
$ I(0,0) = (3/Z^2) \int (d^d l d \omega/(2\pi)^{d+1}) V(l,\omega) 
 G^2 (k_F +l,\omega)$ is non-zero, and $Z$ renormalizes down  from $Z=1$. The same factor $I(0,0)$ 
 appears in the diagrammatic series for 
$\Gamma_c^\omega$. For weakly momentum dependent $V(q, \omega)$, the
 diagrams for $\Gamma_c^\omega$ factorize and  form geometric series.
 In this situation, $\Gamma_c^\omega$  reduces to 
\begin{eqnarray}
\Gamma_c^\omega (\theta) &=& \frac{3}{2} V(\theta) \left( 1 + Z^2 I(0,0)+
Z^4 I^2(0,0) + ...\right) \nonumber \\
&&= \frac{3}{2} V(\theta) ~\frac{1}{1 - Z^2 I(0,0)}
\label{1.36}
\end{eqnarray}
Using the relation (\ref{1.27}) between $Z$ and $I(0,0)$, we find
$1 - Z^2 I(0,0) =Z$, hence
\begin{eqnarray}
&&\Gamma_c^\omega (\theta) = \frac{3}{2Z} V(\theta),~ {\text and} \nonumber \\
&&f_c (\theta) = Z^2 \Gamma_c^\omega (\theta) = \frac{3Z}{2} V(\theta)
\label{1.38}
\end{eqnarray}   
Using (\ref{1.38}), we immediately find that  the diagrammatic expression for the effective mass, Eq. 
(\ref{1.33_b}), is equivalent to the expression for the 
effective mass in the Fermi liquid theory, Eq. (\ref{1.33}).

For strong momentum dependent $V(q, \omega)$, the diagrammatic 
series for $\Gamma_c^\omega (\theta)$ cannot be factorized, and we 
 didn't find how to  prove Eq. (\ref{1.38}) by summing up the diagrams.

\section{The crossover from $\Sigma (k)$ to $\Sigma (\omega)$}

We now return to the results for the effective mass, and the 
 quasiparticle $Z$-factor, Eqs (\ref{1.27}) and (\ref{1.28}).
Assume momentarily that the interaction is purely static. Then $Z=1$
 and 
\begin{equation}
m^* = \frac{m}{1-\lambda}
\label{2.1}
\end{equation}
As we said, when the system approaches the FQCP and $d\leq 3$, 
$\lambda$ increases and eventually becomes larger than $1$. According to 
(\ref{2.1}), at $\lambda=1$, the effective mass diverges, i.e., the
 Fermi-liquid state becomes unstable. This effect was noticed in \cite{khodel,yak_khodel}.
 The authors of \cite{khodel} argued that for
 $\lambda >1$, the Fermi liquid state is replaced by the new state of matter
 which they termed as ``fermionic condensate''. 

The issue we now address is whether Eq. (\ref{2.1}) is valid at $\lambda = O(1)$.  This equation
 was obtained assuming that  $Z=1$.
   We will argue that the renormalization of $Z$ can only be neglected
 at parametrically
 small $\lambda$, while at $\lambda = O(1)$,
 the renormalization of $Z$ 
 is essential. When this renormalization is included, the effective mass remains finite at $\lambda = O(1)$, and diverges only at FQCP, where $\lambda$ becomes infinite. 

\begin{figure}
\includegraphics[width=\FigWidth]{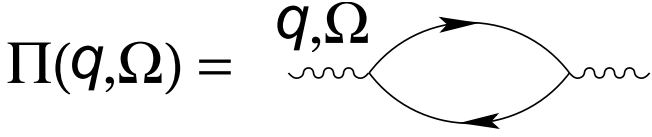}
\caption{\label{fig_4} 
The diagram for the bosonic self-energy $\Pi (q, \Omega)$.}
\end{figure}

To understand this, consider the renormalization of the $Z$ factor in more detail. We remind that it originates from the frequency dependence of the 
 the full vertices and the 
fully renormalized effective interaction $V_{full} (q, \Omega)$. 
The fully renormalized $V_{full} (q, \Omega)$ differs from 
$V (q, \Omega)$ by the 
 bosonic self-energy $\Pi (q,\Omega)$ (see Fig.~\ref{fig_4}):
\begin{equation}
V^{-1}_{full} (q, \Omega) = V^{-1} (q, \Omega) + a^2 \Pi (q, \Omega)
\label{dec_29_1}
\end{equation}
 The contribution 
 to $\Pi (q, \Omega)$ from high-energy fermions is already incorporated into $V(q, \Omega)$ and  must therefore be neglected  to avoid 
double counting.  However, the polarization 
bubble contains also the Landau damping term which comes from internal fermions
 in the bubble with energies smaller than $\Omega$. This term
 is not included into $V(q, \Omega)$ as in the 
the RG procedure  that gives $V(q, \Omega)$ one assumes that 
internal frequencies are much larger than external frequencies. 

We verified a'posteriori that the frequency dependence of the effective interaction is more
 relevant than the frequency dependence of the full vertices 
(the latter leads only to minor corrections at  $\lambda \leq 1$).\
  Accordingly, we
 assume that the full vertices remain static, 
 absorb the vertex renormalization factor $\Gamma^2$ into $g$,
 and focus on $V_{full} (q, \Omega)$.  The frequency dependence of 
 $V_{full}(q,\Omega)$ comes from two sources. First, 
  $V(q,\Omega)$ already  possess some regular
 frequency dependence, see Eq. (\ref{1.21}). Second,
 $\Pi (q, \Omega)$ also introduces frequency dependence into $V_{full} (q, \Omega)$. For small $q$ and even smaller $\Omega < v_F q$, 
\begin{equation}
\Pi (q, \Omega) = \frac{m}{\pi} \left(\frac{|\Omega_m|}{v_F q}\right)
\label{2.7}
\end{equation}
Substituting it into (\ref{dec_29_1}) and using (\ref{1.21}) for $V(q, \Omega)$, we obtain
\begin{equation}
V_{full} (q, \Omega) = \frac{g}{a^2} \frac{1}{q^2 + \xi^{-2} + \frac{\Omega}{q} \gamma + \left(\frac{\Omega}{v_F}\right)^2}
\label{2.8}
\end{equation}
where $\gamma \sim g/(v^2_F a)$, and we used the facts that $v_s \sim v_F$ and 
$ap_F = O(1)$.
At small frequencies and small momenta, the Landau damping term obviously dominates over the regular, $\Omega^2$ term in (\ref{2.8}). We therefore will
 only keep the Landau damping term. 
 
We now use Eq. (\ref{2.8}) to compute 
$I(0,0)$ and $Z$ (see Eqs. (\ref{1.22}) and (\ref{1.27})).
According to Eq. (\ref{1.11}), 
\begin{eqnarray}
I(0,0) &=& -3 \frac{g}{a^2} \int \frac{d^d q d \Omega_m}{(2\pi)^{d+1}}~ \frac{1}{q^2 + \xi^{-2} + \frac{\Omega}{q} \gamma} \nonumber \\
&&\times ~\frac{1}{(\Omega_m + i v_F q \cos \theta)^2}
\label{2.9}
\end{eqnarray}
 Integrating in (\ref{2.9}) first over frequency and then over momentum
  and introducing
\begin{equation}
\alpha = \xi^{-2}/\gamma v_F \sim (a/\xi)^2 (W/g),
\end{equation}
 we obtain 
\begin{equation}
I(0,0) = \lambda_d f(\alpha)
\label{2.10}
\end{equation}
where $f(0) =1$, and at large $\alpha$, $f(\alpha)$ vanishes as some power of
 $\alpha$. In $d=2$, $f(\alpha \ll 1) = 1-0.847 \alpha^{1/2} + ...$ and
 $f(\alpha \gg 1) \approx \log \alpha/(2\alpha)$.

We now analyze the result. That at large $\alpha$, $I(0,0)$ is
 small could be anticipated as $\alpha = \infty$ corresponds to $\gamma =0$, in which case the frequency integral in (\ref{2.9}) vanishes, i.e., 
$I(0,0) =0$. Less anticipated is, however, the fact that 
in the opposite limit of small $\alpha$, $f (\alpha \ll 1) \approx 1$, and 
 $I(0,0) \approx \lambda_d$ becomes independent on $\gamma$. 
 This implies that at small $\alpha$, the regular term in the fermionic self-energy is of the same order as the singular term. Furthermore,
 one can easily verify   that at $\alpha \ll 1$,  
$I(0,0)$ comes from
 fermions in the vicinity of the Fermi surface, just as the
 anomalous part of $I$. Combining regular and anomalous terms in the self-energy, we then obtain
\begin{equation}
\Sigma (k, \omega) = \Sigma_n (k, \omega) + \Sigma_{an} (k) = i \lambda_d ~\omega 
\label{2.11}
\end{equation}
The $k-$dependence of $\Sigma$ is canceled out between $\Sigma_{n} (k, \omega)$ and $\Sigma_{an} (k)$.
We see that the self-energy is totally different at small $\alpha$ and at large $\alpha$. At large $\alpha$, i.e., at some distance away from FQCP, 
the anomalous part of the self-energy well exceeds the regular part, and $\Sigma (k, \omega) \approx \Sigma (k)$. At small $\alpha$, both regular and anomalous parts of the self-energy are of the same order, and the sum of the two yields $\Sigma (k, \omega) \approx \Sigma (\omega)$. In this last case, 
$I(0,0) = \lambda_d$, hence 
\begin{eqnarray}
\frac{1}{Z} &=& 1 + \lambda_d~~{\text and}\nonumber \\
m^* &=& \frac{m}{1 - Z \lambda_d} = m ~(1 + \lambda_d)  
\label{1.28_1}
\end{eqnarray}
 Eq. (\ref{1.28_1})  
implies that $m^*$ does not diverge before FQCP.

The crucial issue is at what $\lambda_d$ the system experiences the crossover
 from $\Sigma (k)$ to $\Sigma (\omega)$. 
If the crossover does not occur up to 
 $\lambda_d = O(1)$, the effective mass
 still diverges at $\lambda_d =1$, and the eventual transformation to 
 $\Sigma (\omega)$ becomes meaningless. However, the crossover  occurs
 already at small $\lambda$, when the calculations are under control. 
 To see this, we note that $\Sigma (k, \omega)$ 
is the scaling function of the single parameter $\alpha$. The crossover in the self-energy then obviously occurs at $\alpha = O(1)$.
Using $\alpha \sim (a/\xi)^2 (W/g)$ and the definition of $\lambda_d$, we obtain that 
\begin{equation}
\alpha^{-1} \sim \lambda_d \left(\frac{\xi}{a}\right)^{d-1}
\label{2.12}
\end{equation}
Hence the crossover from $\Sigma (k)$ to $\Sigma (\omega)$ occurs at
\begin{equation}
\lambda_d \sim \left(\frac{a}{\xi}\right)^{d-1} \sim  \left(\frac{g}{W}\right)^{(d-1)/2}  \ll 1
\label{2.13}
\end{equation}
This is the main result of the paper. We see that the crossover in the self-energy occurs already at small $\lambda_d$, well before the effective mass apparently diverges. This implies that Eq. (\ref{2.1}) is only applicable at small $\lambda$, and it cannot be extrapolated to $\lambda =1$.

\subsection {$\Sigma (\omega)$ as an ``anomaly''}

 At a first glance, the fact that
  $\Sigma_n$ and $\Sigma_{an}$ are comparable above the crossover to $\Sigma (\omega)$, invalidates
 the idea that the self-energy in a Fermi liquid theory can be 
 viewed as an anomaly. It turns out, however, that the self-energy $\Sigma \approx \Sigma (\omega)$ can be also identified with the anomaly, however the
 regularization procedure must be different from the one we used
 to obtain $\Sigma \approx \Sigma (k)$. 
 To understand this, consider again the full self-energy, Eqs. (\ref{1.10}), (\ref{1.11}), but now substitute into (\ref{1.11}) the  full $V_{full} (q, \Omega)$ instead of the bare $V(q, \Omega)$. The anomalous part of the self-energy is associated with the splitting of the double pole and is $\Sigma_{an} =
\lambda_d (i \omega -\delta k) J$, where $J$ is  determined by the integral
\begin{equation}
J = \frac{1}{2\pi} \int \frac{d q_1 d\Omega}{(\omega + \Omega + i v_F (\delta k + q_1))(\Omega + i v_F q_1)}
\label{2.14}
\end{equation}
In Sec.\ref{se}, we evaluated this integral by integrating over 
 frequency first. 
 The  frequency integration restricted the integral over $q_1$ 
to $-\delta k < q_1 <0$ (for positive $\delta k$). The result then was 
\begin{equation}
J = J_k = \frac{\delta k}{i\omega - \delta k} 
\label{2.15}
\end{equation}
and it yielded $\Sigma_{an} = \lambda_d v_F \delta k$, i.e., the self-energy below the crossover. 
 However, as  we already have said,
 the $2D$ integral in (\ref{2.14}) is formally ultraviolet divergent, and  
 the result of the integration over momentum and frequency depends
 on the order of the integration. To check this, let's integrate over $dq_1$ first. The integration proceeds in the same way as before, but now
\begin{equation}
J = - \frac{1}{2\pi v_F} \int d \Omega \int dx \frac{1}{(x + v_F \delta_k - i(\omega + \Omega))(x - i \Omega)}
\label{2.16}
\end{equation}   
where $x = v_F q_1$. The integral over $dx$ is nonzero when the two poles
are located in different half-planes of complex $x$, i.e., when $-\omega < \Omega < 0$ (for $\omega >0$). Performing the integration, we obtain
\begin{equation}
J = J_\omega = \frac{1}{v_F}~\frac{i \omega}{i\omega - \delta k}
\label{2.17}
\end{equation}
Using this result, we obtain $\Sigma_{an} = i \lambda_d \omega$, which coincides with the self-energy above the crossover. 

Comparing $J_k$ and $J_\omega$, we see that they differ by a constant $(= -1/v_F)$. Which of the two results is the correct one? 
The  answer depends on  the functional form of  $V_{full}(q, \Omega)$ 
 which regularizes the integrals  (\ref{2.14}) and (\ref{2.16}). 
Indeed, Eqs. (\ref{2.14}) and (\ref{2.16})
 are valid as long as 
 $V_{full}(q_1, \Omega) = \int d q_{\perp}~
V_{full}(q_1, q_{\perp}, \Omega)$ can be approximated by a constant and absorbed into $\lambda_d$. 
 At small $q_1$ and $\Omega$, this is true if the system is not at FQCP. However, at large
 $q_1$ and $\Omega$, the effective interaction obviously vanishes. 
This implies that once we leave  $V_{full}(q_1, \Omega)$  in 
the integrand for $J$, the $2d$
 integral over $q_1$ and $\Omega$ will become ultraviolet convergent. 
 Then $J$ must be the same independent on what 
integration is performed first. 

Now,  $V_{full}(q_1, \Omega)$ vanishes at large $\Omega$ and at large $q_1$, i.e., both can serve as cutoff. 
 The momentum cutoff at  $q_1 \sim  1/a$ 
 is imposed by the lattice. The frequency cutoff is due to the fact 
 $\Pi (q, \Omega)$ is linear in $\Omega$, hence 
 $V_{full} (q_1, \Omega)$ vanishes at large
 frequencies. For $\alpha \gg 1$, the frequency dependence of $V_{full} (q_1, \Omega)$ is negligible, and the regularization of the integral for $J$ is provided by the momentum cutoff at $q_1 \sim 1/a$. In this situation, it is natural to integrate over 
$\Omega$ first in infinite limits, and then integrate over $q_1$ up to the cutoff. After the integral over $\Omega$ is evaluated, the subsequent
 integration over $q_1$ is  restricted to a much narrower range
 than the cutoff, and we obtain $J = J_k$ independent on the cutoff.
As an exercise, we verified that the same result $J = J_k$ is obtained at $\alpha \gg 1$ by
 integrating first over $q_1$, up to the cutoff, and then integrating over frequency. After the frequency integration, the momentum 
cutoff disappears from $J =J_k$.  
 
For $\alpha \ll 1$, the situation is different as for typical $\Omega \sim v_F q_1 \sim v_F \xi^{-1}$, the Landau damping term in $V_{full} (q_1, \Omega)$
 exceeds the static part of the effective interaction. This implies that the
 regularization is now provided by  the dynamical part of 
 $V_{full} (q_1, \Omega)$.
 In this situation, it is natural to first integrate over $q_1$, in infinite limits, and then integrate over $\Omega$. The result is $J = J_\omega$. Again, as an exercise, we verified that $J = J_\omega$ at $\alpha \ll 1$ 
 is reproduced if we integrate
 over $\Omega$ first, and then integrate over $q_1$. The Landau damping term
 in $V_{full} (q_1, \Omega)$ is relevant at the 
intermediate stages of the calculations, but disappears from the final answer.

This consideration implies that at $\alpha \ll 1$, the self-energy $\Sigma (k, \omega)$ can again be divided into the anomalous and regular parts
$\Sigma (k, \omega) = \Sigma_n (k, \omega) + \Sigma_{an} (k, \omega)$,
 but now
\begin{equation}
\Sigma_{an} (k, \omega) = \Sigma (\omega) = i \lambda_d \omega
\label{2.18}
\end{equation}
and
\begin{equation}
\Sigma_{n} (k, \omega) = (i \omega - v_F \delta k){\tilde J} (0,0)
\label{2.19}
\end{equation} 
where
\begin{equation}
{\tilde J} (0,0) = J(0,0, \alpha) - J(0,0,\alpha =0)
\label{2.20}
\end{equation}    
 is small at $\alpha \ll 1$. In particular, for $d=2$, ${\tilde J} (0,0) \propto \lambda_d \alpha^{1/2} \sim (g/W)^{1/2} \ll 1$.

\section{Strong coupling}

We now consider what happens when the dimensionless coupling $\lambda$
 becomes large, i.e., the system falls into the strong coupling regime.
For definiteness, we focus on $d=2$. 

In general, the strong coupling case can be hardly treated diagrammatically, as 
 one needs to include infinite series of both self-energy and vertex correction diagrams. In our case, the situation, however,
 simplifies as we still have
 $g/W$ as the small parameter. As we just said, in this situation, 
only the anomalous self-energy, which depends  on frequency,
 becomes large when $\lambda_{d=2} = \lambda \geq 1$, whereas the regular
 part of the self-energy remains small in $(g/W)^{1/2}$ even when $\lambda \geq 1$. We verified that in this situation,\\

(i) the quasiparticle density of states
 \begin{equation}
N(\Omega) = -N_0 Im\left[\int \frac{d \epsilon_k} {\pi} \frac{1}{i \Omega + \Sigma (\Omega) - \epsilon_k}\right] = N_0 ~{\text sign}~ \Omega  
\label{dec_30_1}
\end{equation}
 remains the same as for free fermions in the presence of  $\Sigma (\Omega)$,\\  
(ii) vertex corrections at $v_F q \gg \Omega$ 
 remain small in $(g/W)^{1/2}$ even when $\lambda$ becomes large\\

(iii) the corrections to the Landau damping term from various
 insertions into the particle-hole bubble are also small in 
$(g/W)^{1/2}$, i.e., $\Pi (q, \Omega)$ is still determined by Eq. (\ref{2.7}),
 even at strong coupling. 

The  smallness of vertex corrections and corrections to 
$\Pi (q, \Omega)$ is the key element of Eliashberg-type theories~\cite{eliashberg}. 
 If the self-energy depended on $k$, Eq. (\ref{dec_30_1}) 
 would not hold and vertex corrections and corrections to $\Pi (q, \Omega)$ 
 would not be small. 

Without vertex corrections and corrections to (\ref{2.7}), the full self-energy at {\it arbitrary} $\lambda$ above the crossover is still given by Eq. (\ref{1.10}), but with the overall $Z$, with $v^*_F$ instead of $v_F$, and with
 $V_{full} (q, \Omega)$ given by (\ref{2.8}). Evaluating the self-energy
 and neglecting the regular piece in $\Sigma (k, \omega)$, we obtain 
\begin{equation}
\Sigma (k, \omega) = \Sigma_{an} (\omega) = i \omega \lambda \left(Z \frac{m^*}{m}\right)
\label{2.18_a}
\end{equation}
Substituting this result into $G(k, \omega)$, we obtain a set of two coupled equations for $Z$ and $m^*/m$:
\begin{eqnarray}
&&\frac{1}{Z}  = 1 + \frac{m}{m^*} \lambda Z \nonumber \\
&&\frac{1}{m^*} = \frac{1}{m} \left( 1 - \lambda Z\right)
\label{2.19_a}
\end{eqnarray}
These equations are generally different from the weak coupling version,
Eqn. (\ref{1.28_1}) due to the presence of the extra $Z m^*/m$ in the equation for $1/Z$. However, solving   these two equations, we obtain
 the same result as before:
\begin{equation}
m^* = m (1 + \lambda); ~~Z = \frac{1}{1+ \lambda}
\label{2.20_a}
\end{equation}
This equivalence with the weak-coupling result is the consequence of the fact that $~~Z \frac{m^*}{m} =1$ is the invariant of the set (\ref{2.19_a}).

Eq. (\ref{2.20_a}) is another key result of the paper. We see that the effective mass remains finite all the way up to a FQCP, where it diverges together with $\lambda$, and
 Fermi liquid description breaks down. Simultaneously, the quasiparticle $Z$
 factor gradually decreases with increasing $\lambda$ and only 
 vanishes at a FQCP. This implies that 
 the intermediate phase near a FQCP does not develop  both at weak and strong coupling.

\begin{figure}
\includegraphics[width=\FigWidth]{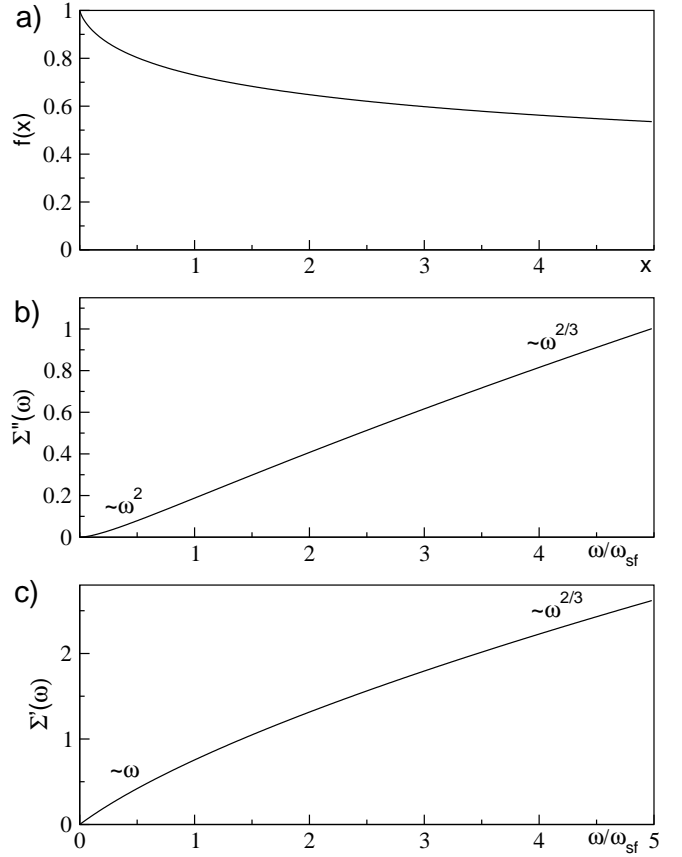}
\caption{\label{fig_5} 
Upper panel. The scaling function $f (x)$ from Eq. \protect\ref{2.29}.
Middle and lower panels: imaginary and real parts of the fermionic self-energy 
 near the ferromagnetic quantum-critical point. Observe that $\Sigma^{\prime \prime}$ is almost linear in $\omega$ in a wide frequency range.}
\end{figure}

For completeness, we also present the result for the fermionic 
 self-energy $\Sigma (k, \omega) = \Sigma (\omega)$ at arbitrary frequencies~\cite{sc_qcp}. 
Substituting $V_{full} (q_1, q_{\perp}, \Omega)$ into the integral for the self-energy and integrating  over $q_1$, we obtain, for positive $\omega$
\begin{equation}
\Sigma (\omega) = i\frac{2}{\pi} ~\lambda~ \int_0^\omega d \Omega
~\int_0^\infty \frac{x dx}{x^3 + x + \Omega/\omega_{sf}}
\label{2.24}
\end{equation}
where $\omega_{sf} = \xi^{-3}/\gamma \sim (W^2/g) (a/\xi)^3$. 
The result can be cast into 
\begin{equation}
\Sigma (\omega) = i ~\lambda~\omega
 f \left(\frac{\omega}{\omega_{sf}}\right)
\label{2.29}
\end{equation}    
The scaling function $f (x)$ is plotted in Fig.\ref{fig_5}
At small $x$ $f(x) = 1 + O(x)$. At large $x$,
\begin{equation}
 f (x \gg 1) \approx \frac{3}{2\pi (x^{1/3})}.
\label{jan_10_1}
\end{equation}
Then, at small $\omega < \omega_{sf}$, we recover the Fermi-liquid behavior, 
 Eqs. (\ref{2.18_a}), (\ref{2.20_a}). 
At $\omega > \omega_{sf}$, the system falls off into the quantum-critical regime, and $\Sigma (\omega) = i \omega^{2/3} \omega_0^{1/3}$, where $\omega_0 \sim \omega_{sf} \lambda^3 \sim g^2/W$ does not depend on $\xi$. Note that still, 
$\omega_0 \ll W$, i.e., quantum-critical behavior is confined to
 low frequencies. At the FQCP, $\omega_{sf}$ vanishes, and $\omega^{2/3}$ behavior extends down to $\omega =0$~\cite{sc_qcp,belitz,aim}.

\begin{figure}
\includegraphics[width=\FigWidth]{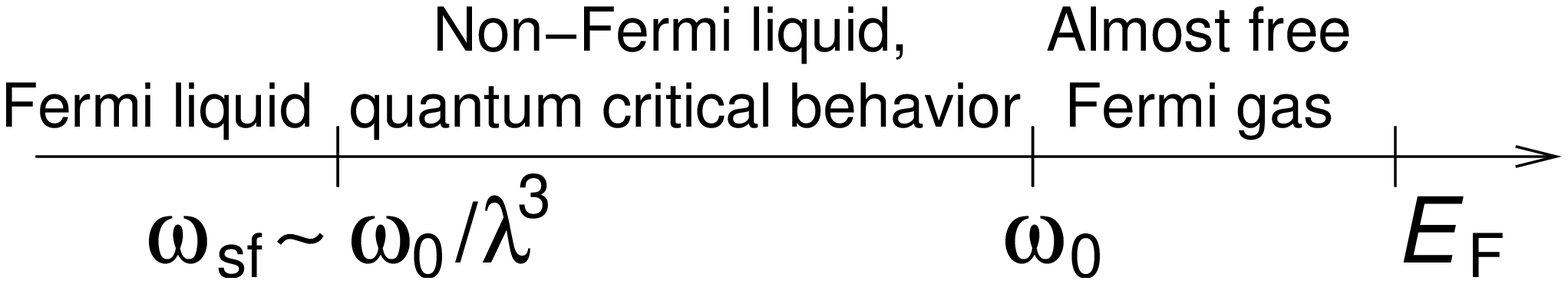}
\caption{\label{fig_6} 
The schematic behavior of the self-energy at various frequencies.} 
\end{figure}

 The system behavior at various frequencies is schematically presented in 
Fig. \ref{fig_6}. At the FQCP, $\omega_{sf}$ vanishes, and $\omega^{2/3}$ behavior extends down to $\omega =0$. 

For arbitrary $3>d>1$, in the quantum-critical regime, $\Sigma (\omega) \propto \omega^{d/3}$. In three dimensions, in the quantum-critical regime, 
$\Sigma (\omega) \propto i \omega \log |\omega|$, i.e., in real frequencies
$Re \Sigma \propto \omega \log|\omega|, ~~Im \Sigma \propto \omega$ 
(a marginal Fermi liquid behavior).  
The quantum-critical behavior in $d =2$ and $d=3$ has been extensively studied recently both theoretically~\cite{sachdev,hf-th,advances,meta,sc_qcp,bedell,belitz}
 and experimentally~\cite{exp,hf-ex,gegen},
 and we refer the reader to the existing literature on this subject.

\section{Conclusion}

To conclude, in this paper we considered a metal near an itinerant 
 ferromagnetic quantum-critical point. We assumed that the residual interaction  $g$
 between fermions and low-energy ferromagnetic collective spin excitations 
 is smaller than the fermionic bandwidth $W \sim E_F$. The
 dimensionless coupling constant $\lambda$ is then small away from a
 ferromagnetic quantum-critical point, at $\xi \sim a$.

We argued that at $d\leq 3$, the dimensionless coupling increases as the system approaches a FQCP,
 and $\lambda$ diverges at a FQCP. We computed fermionic self-energy and found that at weak coupling, $\Sigma (k, \omega) = \Sigma (k)$, and $m^*/m = 1/(1-\lambda)$. If this behavior extended to $\lambda = O(1)$, the effective mass would 
diverge at some distance away from a FQCP. However, we found that already at
 small $\lambda \sim (g/W)^{(d-1)/2}$, the self-energy crosses over from 
$\Sigma (k)$ to $\Sigma (\omega)$, and the result for the effective mass changes to $m^* =m(1+\lambda)$. This implies that the effective mass does not diverge,
 and the Fermi liquid description remains valid everywhere in the paramagnetic phase. 
We argued that the result $m^* =m(1+\lambda)$ remains
 valid  all the way up the FQCP,
 where $m^*$ diverges together with $\lambda$. 

Our consideration is complimentary to the recent analysis of the 
 crossover from $\Sigma (k)$ to $\Sigma (\omega)$ in $2d$ isotropic systems near a density-wave instability at a finite $q_0$~\cite{cgy}. In both cases, the 
crossover occurs at small $\lambda \sim (g/W)^{1/2}$,
 i.e., before the system enters into the
 strong coupling regime.
Still, the finite $q$ case considered in~\cite{cgy} and the $q=0$ case considered here are physically different. In particular,
 the Landau damping term in $\Pi (q, \Omega)$ scales as $\Omega$ at $q \approx q_0$ and as $\Omega/q$ at vanishingly small $q$. Less obvious but
 physically  relevant difference 
 is that  near the transition at $q_0$, the crossover to $\Sigma (\omega)$ occurs at vanishingly small coupling if we set the fermionic bandwidth to infinity~\cite{cgy}. In other words
 the crossover near a QCP at a finite $q_0$ 
 occurs at $\lambda \sim (g/W)^{1/2}$, where $W$ literally has the 
 meaning of the bandwidth.  At the same time,
 near the ferromagnetic transition, the crossover occurs at a {\it finite}
 $\lambda$ even if one sets the bandwidth to infinity 
 and linearizes the fermionic spectrum, as we actually did.
 In this case, $W$  has the meaning of the Fermi energy 
$W = E_F = v_F p_F/2$, rather than of the bandwidth. 

The present analysis does not invalidate completely the idea that the effective mass can diverge before the system reaches a quantum-critical point. We
 proved that this does not happen as long as $g/W$ is small, and the calculations in the crossover regime are under control. What happens at $g \geq W$ is an open issue, and we refrain from speculating what the system behavior may be.
 In any event, for $g$ comparable to the bandwidth, one cannot depart from the spin-fermion model as the very idea that one can integrate out high energy fermions and obtain an effective model for low-energy degrees of freedom becomes 
 meaningless.

It is my pleasure to thank M. I. Kaganov for numerous conversations during the 
 completion of this work. I acknowledge useful conversations 
with V. Galitski, V. Khodel, 
D. Maslov, C. Pepin, and V. Yakovenko. I am thankful to C. Pepin and V. Yakovenko for critical reading of the manuscript and useful comments. 
The work  was supported by NSF 
 DMR-0240238 and the Center for Material Theory at UMD.


\begin{thebibliography}{99}
\bibitem{1}  L.D. Landau, JETP, {\bf 35}, 97 (1958).

\bibitem{2}  A.\ A.\ Abrikosov, L.\ P.\ Gorkov, and I.\ E.\
Dzyaloshinski,
\emph{Methods of quantum field theory in statistical physics}, (Dover
Publications, New York, 1963);
 E. M. Lifshitz and L. P. Pitaevski, \emph{Statistical
Physics}, (Pergamon Press, 1980).

\bibitem{3} N. Ashkroft and D. Mermin, \emph{Solid state Physics}, 
(Thomson Learning, 1976);  A.L. Fetter and D.L. Walecka, 
\emph{Quantum Theory of
Many-Particle Systems}, McGraw-Hill, New York, 1971.


\bibitem{new_Greywall} D.S. Greywall, Phys. Rev. B {\bf 27}, 2747 (1983)
 and references therein.

\bibitem{hertz} J.A. Hertz, \prb {\bf 14}, 1165 (1976);
 A.J. Millis, \prb {\bf 48}, 7183 (1993).

\bibitem{doniach} S. Doniach and S. Engelsberg, \prl {\bf 17}, 750
  (1966).

\bibitem{sachdev} S. Sachdev, {\em Quantum Phase Transitions}, Cambridge University Press 
(Cambridge, 1999).

\bibitem{hf-th}  P. Coleman, Nature \textbf{406}, 508 (2000); P. Coleman 
\textit{et al}, J. Phys. Cond. Matt \textbf{13}, 723 (2001). A. Rosch, Phys.
Rev. Lett. \textbf{82}, 4280 (1999); Q. Si \textit{et al}, cond-mat/0011477.

\bibitem{advances} Ar. Abanov, A. V. Chubukov, and J. Schmalian,
Advances in Physics {\bf 52}, 119 (2003).

\bibitem{meta} A.J. Millis, A.J. Schcofield, G.G. Lonzarich, S.A. Grigera,
Phys. Rev. Lett., {\bf 88}, 217204 (2002).

\bibitem{bedell}  Z. Wang, W. Mao and K. Bedell, Phys. Rev. Lett. \textbf{87}%
, 257001 (2001); K.B. Blagoev, J.R. Engelbrecht and K. Bedell, Phys. Rev.
Lett \textbf{82}, 133 (1999); R. Roussev and A. J. Millis, Phys. Rev. B \textbf{63},
140504 (2001).

\bibitem{sc_qcp}  A. V. Chubukov, A. M. Finkelstein, R. Haslinger, and
  D.~K.~Morr, \prl {\bf 90}, 077002 (2003).

\bibitem{khodel} V. A. Khodel, V.P. Shaginyan and M.V. Zverev, JETP Lett, {\bf 65}, 242 (1997); V. R. Shaginyan, Pis'ma Zh.~Eksp.~Teor.~Fiz. {\bf
    77}, 104 (2003) [JETP Lett. {\bf 77}, 99 (2003)].

\bibitem{yak_khodel}
 V. M. Yakovenko and V. A. Khodel, Pis'ma
Zh.~Eksp.~Teor.~Fiz. {\bf 78}, 850 (2003) [JETP Lett. {\bf 78}, 398
(2003)], see also cond-mat/0308380; 

\bibitem{cgy} A. Chubukov, V. Galitski and V. Yakovenko, Phys. Rev. Lett., 
 2005 to appear. 

\bibitem{belitz} S.A. Brazovskii, Sov. Phys. JETP {\bf 41}, 85 (1975);
 D. Belitz, T. R. Kirkpatrick, and T. Vojta, Phys. Re
v. B
\textbf{55}, 9452 (1997); S. Das Sarma and E. H. Hwang, Phys. Rev. Lett.
\textbf{83}, 164 (1999). 
G. Y. Chitov and A. J. Millis, \prl {\bf 86}, 5337
(2001); %
\prb {\bf 64}, 0544414 (2001);
A. V. Chubukov and D. L. Maslov, Phys. Rev. B \textbf{68, }%
155113 (2003); \emph{ibid. }\textbf{69}, 121102 (2004); A. V. Chubukov. 
 C. Pepin and J. Rech, \prl {\bf 92}, 147003 (2004);
D. Belitz, T.R. Kirkpatrick, and Joerg Rollbuehler, \prl {\bf 93}, 155701 (2004).
\bibitem{lonzarich}  Ph. Monthoux and G.G. Lonzarich,
 Phys. Rev. B {\bf 59}, 14598 (1999).

\bibitem{scal} see, e.g., D.J. Scalapino, E. Loh, and J.E. Hirsch,
 Phys. Rev. B {\bf 34}, 8190 (1986). 

\bibitem{gorkov} M. Dzero, L.P. Gorkov, Phys. Rev. B {\bf 69}, 092501 (2004).

\bibitem{anomaly} S. B. Treiman, R. Jackiw, and D. J. Gross, {\it
    Lectures on Current Algebra and its Applications} (Princeton
  University Press, Princeton, 1972), see also 
B.L. Altshuler et al, Europhys. Lett., {\bf 44}, 401 (1998); R. Haslinger and
  A. V. Chubukov, \prb {\bf 68}, 214508 (2003).

\bibitem{eliashberg}G.M. Eliashberg, Sov. Phys. JETP \textbf{11}, 696 (1960);
D.J. Scalapino in \emph{Superconductivity}, Vol. 1, p. 449, Ed. R. D. Parks,
Dekker Inc. N.Y. 1969; F. Marsiglio and J.P. Carbotte, in `The Physics of
Conventional and Unconventional Superconductors', Eds. K.H. Bennemann and
J.B. Ketterson (Springer-Verlag).

\bibitem{aim}  The $\omega^{2/3}$ behavior has also been found in other systems, by similar reasons:  J.~Polchinski,
Nucl. Phys. B {\bf 422}, 617 (1994); B. L. Altshuler, L.B. Ioffe, and A. J. Millis, Phys. Rev. B 
\textbf{52}, 5563 (1995);  V. Oganesyan, S. A. Kivelson, and E. Fradkin,
Phys. Rev. B {\bf 64}, 195109 (2001); H.Y. Kee and Y.B. Kim, 
J. Phys: Cond. Matter {\bf 16}, 3139 (2004); I. Vekhter and A. Chubukov, \prl {\bf 93}, 016405 (2004).
 
\bibitem{exp} G. R. Stewart,  Rev. Mod. Phys. {\textit 73}, 797 (2001) and
references therein.

\bibitem{hf-ex} S.R. Julian and G. G. Lonzarich, Phys. Rev. B \textbf{55},
8330 (1997); N. D. Mathur \textit{et al}, Nature (London) \textbf{394}, 39
(1998); S.S. Sahena \textit{et al}, Nature \textbf{406}, 587 (2000).

\bibitem{gegen} P. Gegenwart, J. Custers, T. Tayama, 
K. Tenya, C. Geibel, G. Sparn, N. Harrison, P. Kerschl, 
D. Eckert, K.-H. Müller, and F. Steglich, J. Low Temp. Phys. 
{\bf 133}, 3 (2003). 
J. Custers, P. Gegenwart, H. Wilhelm, K. Neumaier, Y. Tokiwa, 
O. Trovarelli, C. Geibel, F. Steglich, C. Pepin, and P. Coleman, 
Nature, {\bf 424}, 524-527 (2003)

\end{thebibliography}
\end{document}